\documentclass[preprint,amsmath,amssymb]{revtex4}
\usepackage{pgfplots}
\usepackage{booktabs}
\usepackage{graphicx}
\usepackage{dcolumn}
\usepackage{bm}
\usepackage[skip=2pt,font=footnotesize]{caption}
\usepackage{amsmath}
\usepackage{mathrsfs}
\usepackage[english]{babel}
\usepackage{float}
\usepackage{amsfonts}
\usepackage{amssymb}
\usepackage{dsfont}
\usepackage{bbm}
\usepackage[utf8]{inputenc}
\usepackage[overload]{empheq}
\usepackage{cases}

\begin{document}
\normalsize
\title{Quantum vacuum under mixed boundary conditions: the case for curved spacetime\\}
\author{ Borzoo Nazari  \footnote{borzoo.nazari@ut.ac.ir}
 }
\affiliation{School of Surveying and Geospatial Engineering, College of Engineering, University of Tehran, Tehran, Iran}

\begin{abstract}
Influence of gravity on the quantum vacuum of a massless minimally coupled scalar field under Robin boundary conditions on parallel plates is investigated. We introduce the detailed calculation of the volume energy for the case the gravitational background is weak in its most general form for a static spacetime. It founds that the quantum vacuum usually reacts to the gravitational field by decreasing the Casimir energy. In addition, we find sufficient conditions under which the Casimir force increases. Interestingly, the first order perturbation corrections, are present in the obtained formula for the volume energy. We show that for some specific choices of parameters, the energy is independent of Robin coefficients. Consistency with the literature is shown in some limiting cases and well-known examples are presented for both an increase or decrease in the volume energy.
\end{abstract}
\maketitle
\section{Introduction}
Casimir effect is a typical property of a quantum field under boundary conditions. The boundary may be either a material boundary or a periodic/anti-periodic condition imposed on the spacetime manifold \cite{Asoreya,Myers}. In both cases, the Robin boundary condition (Robin BC), as a generalization of Neumann and Dirichlet conditions, is of great importance. Dirichlet BC requires the wave function to take specific values on boundary while for Neumann BC the derivative of the wave function is specified on boundary. The Robin BC is a third type of condition for which a linear combination of the wave function and its derivative is specified on the boundary. Most traditionally, in physics, we need the three types to vanish on boundary.

For material boundaries, the Robin coefficients play the role of skin depth of the surface \cite{BorzooAnnDerPhys}. 
Without having a physical boundary, a periodic boundary condition on a quantum field may also occur when the spacetime admits some compact dimensions \cite{Asoreya,Myers}. This point shows the importance of the Casimir effect in higher dimensional theories of physics. Extensive studies have been devoted to the Casimir effect in extra dimensions, i.e. in the brane world paradigm, the Randall-Sundrum, the Kaluza-Klein and some other models \cite{Teo2}. In the context of the spacetimes with extra dimensions, the Robin conditions arises naturally for Randall-Sundrum models and some models of quantum gravity \cite{Elizalde},\cite{Teo}. Robin BCs are important also from the point of view of the fact that they are, in contrast to Dirichlet and Neumann ones, can be adopted to be invariant under conformal transformations, see \cite{Saharian}.

The vacuum energy has possible gravitational and cosmological implications in the sense of the well-known question that whether the quantum vacuum gravitates or not. Some authors studied Casimir effect in curved spacetime seeking an answer to this controversial question. \cite{MiltonSaharian,Caldwell,Mahajan,Garattini}.
Calculations of energy-momentum tensor and the Casimir force and pressure in curved spacetime are typically lengthy and cumbersome. There are two approaches for such calculations. Several authors use the standard approach to find mode functions of the corresponding field theory and calculate the Casimir energy directly \cite{Saharian,Sorge,SorgeNew,BorzooEPJC,Bezerra,Muniz1,Muniz2,NouriNazari2}. Many others employ the Green function method along with some standard regularization technique, mostly the covariant point splitting method, to find the energy-momentum tensor and total energy\cite{Bimonte,Bimonte1,Esposito,Napolitano,Milton,Geyer,Dowker}.\\
To be more clear, lets assume the background spacetime as
\begin{eqnarray}\label{eq01}
ds^2 = (1 +2\gamma_0 +2 \lambda_0 z) dt^2- (1+2\gamma_1 +2 \lambda_1 z)\left(dx^2+dy^2+dz^2\right),
\end{eqnarray}
in which $\lambda_0,\lambda_1,\gamma_0,\gamma_1<<1$. 
Using the Green function method, in a series of papers, Bimonte et al \cite{Bimonte,Bimonte1}, Esposito et al \cite{Esposito} and Napolitano et al \cite{Napolitano} analysed the problem of the Casimir plates in Fermi coordinates and calculated the energy-momentum tensor. By Fermi coordinates we mean $\lambda_0\neq0,\gamma_0=\gamma_1=\lambda_1=0$. They found corrections up to $O(\lambda_0)$ compared with the usual Casimir energy in flat spacetime .

In another approach, Sorge \cite{Sorge} has found the total Casimir energy of two parallel plates in the Schwarzschild spacetime up to second order perturbations, i.e. $O(\lambda_0^2)$.
Through a different context, the author and coworkers calculated the total energy of Casimir plates for the case of Fermi coordinates \cite{NouriNazari}. The result was consistent to ones previously found in  \cite{Bimonte,Bimonte1,Esposito,Napolitano} where the Green' function method used. Later, some authors \cite{Bezerra,Muniz1,Muniz2} employed the same method to find corrections to the Casimir energy in the weak field limit of various spacetimes. \\
The total energy as well as the energy-momentum tensor of the Casimir plates in the spacetime (\ref{eq01}) was analysed by the author \cite{BorzooEPJC} for the case of the Neuman and Dirichlet boundary conditions. This paper aims to extend the calculations to Robin boundary condition and correct some misleading calculations in the literature. In connection with this, an important point is worth noting. It is well known \cite{Saharian, saharian5} that the total Casimir energy of plates ,and in general for any quantum field under external boundary condition, consists of a volume energy and a surface energy resides on the boundaries . Although the surface energy vanishes for Neumann and Dirichlet conditions on boundary, it survives for Robin one's. In \cite{BorzooAnnDerPhys} we found the total energy. Here, we try to compute the volume energy. Moreover, Sorge \cite{Sorge2019} recomputed recently the energy corrections using the different approach of Schwinger action principle for Schwarzschild spacetime. Our results confirm him in a limiting case. 
Calculations of the Casimir energy in curved spacetime usually encounter summation over an expression like
\begin{eqnarray}\label{eq02}
E=\frac{\hbar}{2} \int \sum_{\omega_n,k} f(\omega_n,k) d^2k,
\end{eqnarray}
where $\omega_n$ satisfies another equation
\begin{eqnarray}\label{eq03}
g(\omega_n)=0,
\end{eqnarray}
which stems from imposition of boundary conditions. Except for some special cases, i.e. in flat spacetime and some highly symmetric spacetimes, obtaining a closed and explicit form for series (\ref{eq02}) under condition (\ref{eq03}) is often impossible. Although some improved calculations have been developed in curved spacetime \cite{Sahariannew, Teo,Teo2}, they still consist of uncomputable sums and integrations. Our plan is to compute the energy explicitly up to second order perturbation for the case of the scalar field under Robin conditions in the curved background described by (\ref{eq01}).

We see that the pertubative method is more complicated than one might suspect at the begining and simple calculations done in the literature are insufficient even in the much more convenient case of the problem under Dirichlet and Neumann boundary conditions. As the Robin boundary condition is more general than Dirichlet and Neumann types, the detailed calculations we present here will clear the situation for all special cases studied by others.

The structure of the paper is as follows. In section II, we find the wave function in the space between the plates. In section III, using Robin conditions on plates, the characteristic equation (\ref{eq03}) is realized. Section IV is devoted to the calculations of the energy under condition (\ref{eq03}). In section V, some examples are provided. The Conclusion is the final section.

\section{The approximated wave function}
Suppose there are two parallel plates separated by a distance $l$ and placed perpendicular to the radial direction in distance $R$ from the center of some gravitational source as depicted in Fig.1.
Assume $m\ll R,\; l<R$ where $m=\frac{GM}{c^2}$ is the gravitational mass. These conditions are equivalent to $\lambda_0,\lambda_1,\gamma_0,\gamma_1<<1$.
\begin{figure}
\begin{minipage}[c]{0.8\linewidth}
\includegraphics[width=0.5\linewidth]{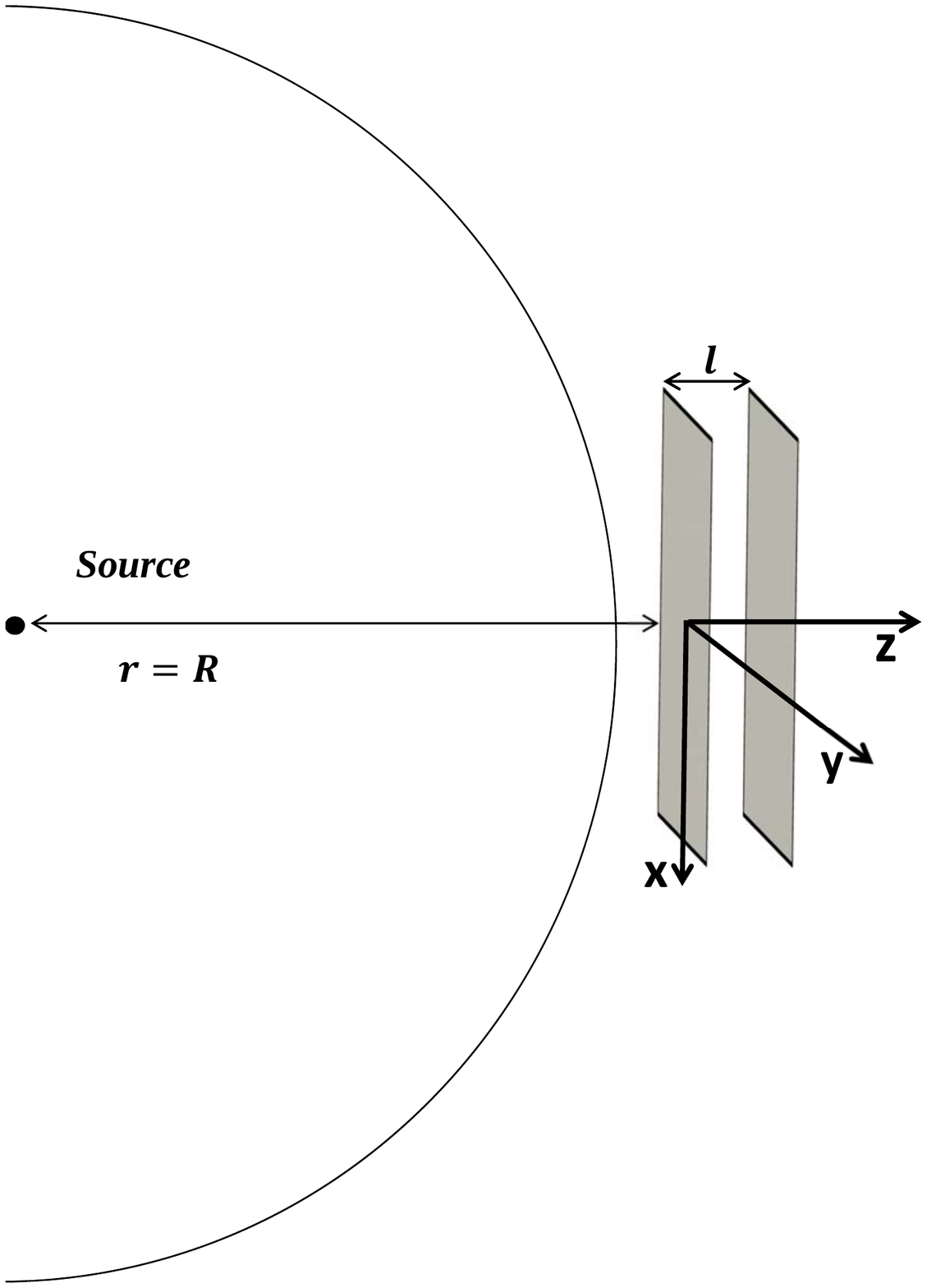}\label{fig:apparatus}
   \caption{Two parallel plates at distance $l$ from each other placed in a gravitational field. The metric can be expanded around the point $r=R$ for small $z<R$ \cite{BorzooEPJC}. Deviation of the metric relative to flat space is small in such away that we can assume the metric form (\ref{eq01}) in the space between the plates}
\end{minipage}
\end{figure}
As Fig.1 shows, the rectangular coordinates $(x,y,z)$ have been introduced.
It is shown that any static \cite{BorzooEPJC}, and under specific conditions any stationary \cite{SorgeNew}, spacetime can be expanded in such away that the metric takes the form (\ref{eq01}). 
The massless scalar Klein-Gordon equation $\Box \Phi=0$ is
\begin{eqnarray}\label{eq04}
g^{00}\partial_t^2\Phi+g^{33}(\partial_x^2\Phi+\partial_y^2\Phi)+\frac{1}{\sqrt{-g}}\partial_z(\sqrt{-g}g^{33}\partial_z\Phi)=0.
\end{eqnarray}
From now on, we calculate everything up to second order perturbation in terms of the parameters $\gamma_0, \lambda_0, \gamma_1, \lambda_1 <<1$. With the ansatz
\begin{eqnarray}\label{eq05}
\Phi(x) = C e^{-i\omega t} e^{ik_x x}e^{ik_y y}Z(z),
\end{eqnarray}
where $C$ is the normalization constant, the Klein-Gordon equation reads
\begin{eqnarray}\label{eq06}
Z''(z)+\partial_z\ln{|\sqrt{-g}g^{33}|} Z'(z)+(-g^{00}g_{33}\omega^2-k_{\perp}^2)Z(z)=0,
\end{eqnarray}
in which $k_\perp^2=k_x^2+k_y^2$. Expanding the metric components as
\begin{eqnarray}\label{eq07}
\begin{split}
&g^{33}\sqrt{-g}=\sqrt{g_{00}|g_{33}|}=1+\gamma_1+\gamma_0+(\lambda_1+\lambda_0)z+O(\lambda^2,\gamma^2),\\ &\ln{|\sqrt{-g}g^{33}|}=\gamma_1+\gamma_0+(\lambda_1+\lambda_0)z+O(\lambda^2,\gamma^2)\\
&\partial_z\ln{|\sqrt{-g}g^{33}|}=(\lambda_1+\lambda_0)+O(\lambda^2,\gamma^2)\\
&-g^{00}g_{33}=1-2(\gamma_0-\gamma_1)-2(\lambda_0-\lambda_1)z+O(\lambda^2,\gamma^2),
\end{split}
\end{eqnarray}
we have
\begin{eqnarray}\label{eq08}
Z''(z)+\lambda Z'(z)+(b+az)Z(z)=0,
\end{eqnarray}
where the prim ($'$) denotes differentiation with respect to $z$ and $a,b$ and $\lambda$ are defined as
\begin{eqnarray}\label{eq09}
a=-2B\omega^2, \;\; b=(1-2A)\omega^2-k_\perp^2,\;\; \lambda\equiv \lambda_1+\lambda_0,
\end{eqnarray}
in which
\begin{eqnarray}\label{eq10}
A\equiv \gamma_0-\gamma_1, \;\; B\equiv \lambda_0-\lambda_1.
\end{eqnarray}
The transformation
\begin{eqnarray}\label{eq11}
W(z)=(1+\frac{\lambda}{2}z)Z(z),
\end{eqnarray}
and the variable change $u=a^{\frac{-3}{2}}(b+az)$ recast (\ref{eq08}) into
\begin{eqnarray}\label{eq12}
W''(u)+uW(u)=0.
\end{eqnarray}
This is the Airy differential equation with the solutions (see the Introduction in \cite{Gradshtyn})
\begin{eqnarray}\label{eq13}
\begin{split}
W(t)=&c_1Ai(-u)+c_2B(-u)\\
&=\sqrt{u}\left(c_3J_{\frac{1}{3}}(\frac{2}{3}u^{\frac{3}{2}})+c_4J_{-\frac{1}{3}}(\frac{2}{3}u^{\frac{3}{2}}) \right).
\end{split}
\end{eqnarray}
Since $a<<1$, i.e. $u>>1$, the asymptotic expansion of the Bessel functions \cite{Gradshtyn}
\begin{eqnarray}\label{eq14}
J_{\frac{1}{3}}(x)\propto\frac{1}{\sqrt{x}}\sin(x+\alpha_0), \; J_{-\frac{1}{3}}(x)\propto\frac{1}{\sqrt{x}}\sin(x+\alpha_1),
\end{eqnarray}
can be used to find
\begin{eqnarray}\label{eq15}
\begin{split}
Z(z)=&D_0(1-\frac{\lambda}{2}z)(1-\frac{a}{4b}z)\sin\left(\frac{2}{3a}(b+az)^{\frac{3}{2}}+\phi_0 \right),\\
=&D_0\left(1-(\frac{\lambda}{2}+\frac{a}{4b})z\right)\sin\left(\sqrt{b}z(1+\frac{a}{4b}z)+\Theta_0 \right),
\end{split}
\end{eqnarray}
where $\phi_0,\; \Theta_0$ and $D_0$ are constants. $D_0$ can be absorbed into $C$ in (\ref{eq05}). 
The constant phase $\Theta_0$ will be found using boundary conditions.

\section{The characteristic equation for frequencies}
The Robin boundary conditions are supposed to be
\begin{eqnarray}\label{eq16}
 \frac{\partial Z(z)}{\partial z}|_{z=0}=\kappa_1 Z(z)|_{z=0}, \\
 \frac{\partial Z(z)}{\partial z}|_{z=l}=-\kappa_2 Z(z)|_{z=l}, \nonumber
\end{eqnarray}
where $\kappa_1$ and $\kappa_2$ are constants. They are taken with opposite signs for seeking simplicity of comparison with the literature in flat spacetime \cite{Saharian}. Equation (\ref{eq16}) will result in

\begin{subequations}
\begin{align}
cot(\Theta_0)&=\frac{1}{\sqrt{b}}(\kappa_1+\frac{\lambda}{2}+\frac{a}{4b}), \label{eq17a}\\
cot(\sqrt{b}l+\frac{a}{4\sqrt{b}}l^2+\Theta_0)&=\frac{1}{\sqrt{b}}\left(\frac{\lambda}{2}+\frac{a}{4b}-\kappa_2(1-\frac{a}{2b}l)\right).  \label{eq17b}
\end{align}
By removing $\Theta_0$ we arrive at
\begin{align}
\tan(\sqrt{b}l+\frac{a}{4\sqrt{b}}l^2)&=\frac{1}{\sqrt{b}}\frac{\kappa_1+\kappa_2(1-\frac{al}{2b})}{1-\kappa_1\kappa_2\frac{1}{b}(1-\frac{al}{2b})+\frac{1}{b}(\frac{\lambda}{2}+\frac{a}{2b})(\kappa_1-\kappa_2)} \label{eq17c},
\end{align}
Note that to eliminate $\Theta_0$, we have used the identity
\begin{align}
\tan(\alpha-\beta)&=\frac{\cot(\beta)-\cot(\alpha)}{1+\cot(\alpha)\cot(\beta)},            \label{eq17d}
\end{align}
up to second order perturbation in terms of $\lambda$ along with $\alpha=\sqrt{b}l+\frac{a}{4\sqrt{b}}l^2+\Theta_0$ and $\beta=\Theta_0$.
\end{subequations}

Equation (\ref{eq17c}) corresponds to (\ref{eq03}). In the following we find $\omega_n$ for important cases $\kappa_1 l,\kappa_2 l>>1$ and $\kappa_1 l,\kappa_2 l<<1$.

\subsection{$\omega_n$ for $\kappa_1 l,\kappa_2 l<<1$}
In this case, the product $\kappa_1\kappa_2$ is negligible and
\begin{eqnarray}\label{eq18}
\tan(\sqrt{b}l+\frac{a}{4\sqrt{b}}l^2)=\frac{1}{\sqrt{b}}(\kappa_1+\kappa_2(1-\frac{al}{2b}))+O(\kappa_i^2)<<1,
\end{eqnarray}
which is equivalent to
\begin{eqnarray}\label{eq19}
\sqrt{b}l+\frac{a}{4\sqrt{b}}l^2=\frac{1}{\sqrt{b}}(\kappa_1+\kappa_2(1-\frac{al}{2b}))+n\pi, \;\;\;\; n=0,1,2,...\; .
\end{eqnarray}
By squaring both sides we find after some steps
\begin{eqnarray}\label{eq20}
b+\frac{a l}{2}-\frac{2}{l}\left(\kappa_2(1-\frac{al}{2b})+\kappa_1 (1+\frac{al}{2b}) \right)=(\frac{n\pi}{l})^2 , \;\;\;\; n=0,1,2,...\; ,
\end{eqnarray}
in which we kept only terms up to $O(\kappa_i^2)$. After some calculations, and making use of equation (\ref{eq09}), it founds that
\begin{eqnarray}\label{eq21}
\omega^2=(1+2A+B l)\left[\omega_0^2+2T_0 \right]+B\frac{\omega^2}{(1-2A)\omega^2-k_\perp^2}(\kappa_2-\kappa_1),
\end{eqnarray}
where $\omega_0^2=k_\perp^2+(\frac{n\pi}{l})^2$ is the corresponding mode frequencies in flat spacetime and $T_0=(\kappa_1+\kappa_2)/l$. The following approximations are frequently used for some function $H$:
\begin{subequations}
    \begin{align}
    &\frac{O(\lambda_i,\gamma_i)}{H(\omega,k_\perp,..)+O(\lambda_i,\gamma_i)}=\frac{O(\lambda_i,\gamma_i)}{H}\left(1-\frac{O(\lambda_i,\gamma_i)}{H}\right)
    =\frac{O(\lambda_i,\gamma_i)}{H}+O(\lambda_i^2,\gamma_i^2),\label{eq22a}\\
    &\sqrt{H+O(\lambda_i,\lambda_i)}=\sqrt{H}+\frac{O(\lambda_i,\lambda_i)}{2\sqrt{H}}+O(\lambda_i^2,\gamma_i^2). \label{eq22b}
    \end{align}
\end{subequations}

The second term in (\ref{eq21}) is an example of the application of (\ref{eq22a}). Equation (\ref{eq21}) is quadratic in $y=\omega^2$ so that we can write
\begin{eqnarray}\label{eq23}
\begin{split}
y^2-My+N&=0,\\
M&=k_\perp^2+(1+2A+B l)\left[\omega_0^2+2T_0 \right]+B(\kappa_2-\kappa_1),\\
N&=k_\perp^2\left((1+2A+B l)\left[\omega_0^2+2T_0 \right] \right).
\end{split}
\end{eqnarray}
Thus, after a calculation, one finds for $\omega^2$
\begin{eqnarray}\label{eq24}
\begin{split}
\omega^2&=M-k_\perp^2+\frac{B k_\perp^2}{M-2 k_\perp^2}(\kappa_2-\kappa_1)\\
&=(1+2A+B l)\left[\omega_0^2+2T_0 \right] \left(1+B\frac{\kappa_2-\kappa_1}{(\frac{n\pi}{l})^2+2T_0}\right).
\end{split}
\end{eqnarray}
This result can also be confirmed using the method of successive approximations.

\subsection{$\omega_n$ for $\kappa_1 l,\kappa_2 l>>1$}
Here, $\kappa_1 \kappa_2$  is the dominant term and (\ref{eq17c}) can be written as
\begin{eqnarray}\label{eq25}
\tan(\sqrt{b}l+\frac{a}{4\sqrt{b}}l^2)=-\sqrt{b}[\beta_1+\beta_2(1+\frac{al}{2b})]+O(\kappa_i^2)+O(\lambda_i^2)<<1,
\end{eqnarray}
which in turn gives
\begin{eqnarray}\label{eq26}
\sqrt{b}l+\frac{a}{4\sqrt{b}}l^2=-\sqrt{b}(\beta_1+\beta_2(1+\frac{al}{2b}))+n\pi,\;\;\;  n=0,1,2,... \; ,
\end{eqnarray}
where $\beta_1=\kappa_1^{-1},\beta_2=\kappa_2^{-1}$. After displacement of some terms and squaring both sides in (\ref{eq26}) we arrive at
\begin{eqnarray}\label{eq27}
b(1+2T_0^\prime)+\frac{a l}{2}+\frac{a}{2}(\beta_1+3\beta_2)=(\frac{n\pi}{l})^2,
\end{eqnarray}
 in which $T_0^\prime=(\beta_1+\beta_2)/l$. Using (\ref{eq09}) again, along with the assumptions ${\omega_0^\prime}^2=(\frac{n\pi}{l^\prime})^2+k_\perp^2$ and $l^\prime=(1+T_0^\prime)l$, the mode frequencies will be found as
\begin{eqnarray}\label{eq28}
\omega=\left(1+A+\frac{B}{2}l-\frac{B}{2}(\beta_1-\beta_2) \right) \omega_0^\prime.
\end{eqnarray}

Note that $l$ is the coordinate distance between the plates which is related to the proper distance by
\begin{eqnarray}\label{eq29}
l_p=\int_0^l \sqrt{-g_{33}}dz=l(1+\gamma_1+\frac{1}{2}\lambda_1l).
\end{eqnarray}
In section IV, we express the final results in terms of $l_p$.

\subsection{some important relations }
Three important terms $sin(2\sqrt{b})$, $sin(2\Theta_0)$ and $cos(2\Theta_0)$ will appear in the expression for the energy in subsequent sections. We find them here for both of the cases mentioned in subsections A and B.
\subsubsection{$\kappa_1 l,\kappa_2 l<<1$}
One can find from (\ref{eq19})
\begin{eqnarray}\label{eq30}
\begin{split}
2\sqrt{b}l&=-\frac{a l^2}{2\sqrt{b}}+D+2n\pi, \;\;\; n=0,1,2,..\\
&\sin{2\sqrt{b}l}=-\frac{a l^2}{2\sqrt{b}}+D, \;\;\; \cos{2\sqrt{b}l}=1,
\end{split}
\end{eqnarray}
in which
\begin{eqnarray}\label{eq31}
D=\frac{2}{\sqrt{b}}(\kappa_1+\kappa_2(1-\frac{al}{2b})).
\end{eqnarray}
Moreover, equation (\ref{eq17a}) shows that $cot(2\Theta_0)<<1$. Therefore, we find
\begin{eqnarray}\label{eq32}
sin(2\Theta_0)=\frac{2}{\sqrt{b}}(\kappa_1+\frac{\lambda}{2}+\frac{a}{4b}),
\end{eqnarray}
which in turn implies
\begin{eqnarray}\label{eq33}
cos(2\Theta_0)=-1+\frac{2}{b}\kappa_1(\lambda+\frac{a}{2b}).
\end{eqnarray}

\subsubsection{$\beta_1 ,\beta_2 <<l$}
In a similar way, we have from (\ref{eq26})
\begin{eqnarray}\label{eq34}
\begin{split}
2\sqrt{b}l&=-\frac{a l^2}{2\sqrt{b}}+D^\prime+2n\pi, \;\;\; n=0,1,2,..\\
&\sin{2\sqrt{b} l}=-\frac{a l^2}{2\sqrt{b}}+D^\prime, \;\;\; \cos{2\sqrt{b}l}=1,
\end{split}
\end{eqnarray}
in which
\begin{eqnarray}\label{eq35}
D^\prime=-2\sqrt{b}(\beta_1+\beta_2(1+\frac{a l}{2b})).
\end{eqnarray}
Putting $\kappa_1=\frac{1}{\beta_1}$ into (\ref{eq17a}) one finds for $\Theta_0$
\begin{subequations}\label{eq36}
\begin{align}
sin(2\Theta_0)&=2\sqrt{b}\beta_1,\label{eq36a}\\
cos(2\Theta_0)&=1. \label{eq36b}
\end{align}
\end{subequations}
In this section, our focus was on simplification of characteristic equation (\ref{eq17c}) as a relization of (\ref{eq03}). The results, i.e. equations (\ref{eq24}), (\ref{eq28}) and (\ref{eq30})-(\ref{eq36}), will be used to find the energy in the next section. This corresponds to imposing (\ref{eq02}) on (\ref{eq03}).

\section{the volume energy}
The classical energy-momentum tensor for a real scalar field is
\begin{eqnarray}\label{eq37}
T_{\mu\nu} &=\partial_\mu\phi\partial_\nu\phi-\frac{1}{2}g_{\mu\nu}g^{\lambda\theta}\partial_\lambda\phi\partial_\theta\phi,
\end{eqnarray}
and the vacuum expectation value of the energy momentum tensor given by
\begin{eqnarray}\label{eq38}
<0|T_{\mu\nu}|0>=\sum_{\mathbf{k}}T_{\mu\nu}[\phi_{\mathbf{k}},\phi^*_{\mathbf{k}}],
\end{eqnarray}
where $T_{\mu\nu}[\phi_{\mathbf{k}},\phi^*_{\mathbf{k}}]$ defined by the bilinear form \cite{Birrell}:
\begin{eqnarray}\label{eq39}
\begin{split}
T_{\mu\nu}[\phi_{\mathbf{k}},\phi^*_{\mathbf{k}}] =\partial_\mu\phi_{\mathbf{k}}\partial_\nu\phi^*_{\mathbf{k}}-\frac{1}{2}g_{\mu\nu}g^{\lambda\theta} \partial_\lambda\phi_{\mathbf{k}}\partial_\theta\phi^*_{\mathbf{k}}.
\end{split}
\end{eqnarray}
Using the wave function (\ref{eq05}) it founds that
\begin{eqnarray}\label{eq40}
\begin{split}
<0|T_{00}^{\phi}|0> &=\frac{1}{2}\sum_\omega\int d^2k_\perp C^2 \left\{\left(\omega^2-\frac{g_{00}}{g_{33}}k^{2}_\perp \right)Z(z)^2 -\frac{g_{00}}{g_{33}}\partial_zZ(z)^2\right\}.
\end{split}
\end{eqnarray}
The volume energy in a static spacetime admitting some killing vector $\zeta^\mu$ is as follows \cite{saharian5}
\begin{eqnarray}\label{eq41}
\begin{split}
E=&\int <0|T^0_{\;\;\nu}|0> \zeta^\nu \sqrt{-g} d^3x\\
=&\frac{A}{2}\sum_\omega \int d^2k_\perp C^2 \left\{\omega^2\int_0^l |g_{33}|\sqrt{\frac{g_{33}}{g_{00}}} Z^2+
k^{2}_\perp \int_0^l \sqrt{-g_{33}g_{00}} Z^2 +\int_0^l \sqrt{-g_{33}g_{00}} \partial_zZ^2\right\}dz,
\end{split}
\end{eqnarray}
where we have used $\zeta^\nu=\delta^\nu_{\;\;0}$ as the Killing vector of a static spacetime and $C=C(\omega,a,b)$. As noted in section (I), the total energy consists of surface and volume energies. None of them are conserved quantities individually. It can be proved that for static spacetime the total energy is conserved \cite{saharian5}.

By defining $\Theta=\sqrt{b}z+\frac{a}{4\sqrt{b}}z^2+\Theta_0$ and using the expansions
\begin{eqnarray}\label{eq42}
\begin{split}
|g_{33}|\sqrt{\frac{g_{33}}{g_{00}}}&=1+3\gamma_1-\gamma_0+(3\lambda_1-\lambda_0)z,\\
\sqrt{-g_{33}g_{00}}=&1+\gamma_1+\gamma_0+(\lambda_1+\lambda_0)z,\\
Z^2=&\frac{1}{2}\left[1-(\lambda+\frac{a}{2b})\right](1-\cos{2\Theta}),\\
\partial_z Z^2=&\frac{b}{2}\left[1-(\lambda-\frac{a}{2b})\right](1+\cos{2\Theta})-\sqrt{b}\frac{1}{2}(\lambda+\frac{a}{2b})\sin{2\Theta},
\end{split}
\end{eqnarray}
it founds that
\begin{eqnarray}\label{eq43}
\begin{split}
I_1=&\int_0^l |g_{33}|\sqrt{\frac{g_{33}}{g_{00}}} Z^2 dz=\frac{l}{2}\left[1+3\gamma_1-\gamma_0+(\lambda_1-\lambda_0-\frac{a}{4b})l\right]-\frac{1}{2}(1+3\gamma_1-\gamma_0)\int_0^l \cos(2\Theta)\\
&\;\;\;\;\;\;\;\;\;\;\;\;\;\;\;\;\;\;\;\;\;\;\;\;\;\;\;\;\;\;\;\;-(\lambda_1-\lambda_0-\frac{a}{4b})\int z\cos(2\Theta),\\
I_2=&\int_0^l \sqrt{-g_{33}g_{00}}Z^2 dz=\frac{l}{2}\left[1+\gamma_0+\gamma_1-\frac{a}{4b}l\right]-\frac{1}{2}(1+\gamma_1+\gamma_0)\int_0^l \cos(2\Theta)\\
&\;\;\;\;\;\;\;\;\;\;\;\;\;\;\;\;\;\;\;\;\;\;\;\;\;\;+\frac{a}{4b}\int_0^l z\cos(2\Theta),\\
I_3=&\int_0^1 \sqrt{-g_{33}g_{00}} \partial_z Z^2dz= \frac{b}{2}\left[1+\gamma_0+\gamma_1+\frac{a}{4b}l\right]l+\frac{b}{2}(1+\gamma_1+\gamma_0)\int_0^l \cos(2\Theta)\\
&\;\;\;\;\;\;\;\;\;\;\;\;\;\;\;\;\;\;\;\;\;\;\;\;\;\;\;\;\;\;\;\;\;\;\;+\frac{a}{4}\int_0^l z\cos(2\Theta) -\frac{\sqrt{b}}{2}(\lambda+\frac{a}{2b})\int_0^l\sin{2\Theta}.\\
\end{split}
\end{eqnarray}
Therefore, the energy reads
\begin{eqnarray}\label{eq44}
\begin{split}
E&=\frac{A}{2}\sum_\omega \int d^2k_\perp C^2 \left\{\omega^2 I_1+
k^{2}_\perp I_2 +I_3\right\},\\
&=\frac{A}{2}\sum_\omega \int d^2k_\perp C^2 \left\{ (I_1+(1-2A)I_2)\omega^2 + I_3-bI_2\right\},
\end{split}
\end{eqnarray}
where we have substituted $k_\perp^2$ from (\ref{eq09}). Furthermore, a computation shows
\begin{eqnarray}\label{eq45}
\begin{split}
I_1+(1-2A)I_2=
&\left(1-\gamma_0+3\gamma_1+\frac{1}{2}(\lambda_1-\lambda_0-\frac{a}{2b})l\right)l\\
&-(1+\gamma_0+3\gamma_1)P_1-(\lambda_1-\lambda_0-\frac{a}{2b})P_2,\\
I_3-bI_2=&
\frac{a l^2}{4}+b(1+\gamma_0+\gamma_1)P_1-\frac{\sqrt{b}}{2}(\lambda+\frac{a}{2b})P_3,
\end{split}
\end{eqnarray}
in which $P_1=\int_0^l \cos(2\Theta)$, $P_2=\int_0^l z\cos(2\Theta)$ and $P_3\int_0^l \sin(2\Theta)$.

\subsection{The parameter $C^2$}
The inner product is defined by \cite{ParkerBook}
\begin{eqnarray}\label{eq46}
(\Phi_1 , \Phi_2) =
-i\int_{\Sigma}\Phi_1(x){\overleftrightarrow{\partial}}_{\mu}
\Phi_2^*(x)[-\mathfrak{g}_{\Sigma}(x)]^{\frac{1}{2}}n^{\mu} d\Sigma,
\end{eqnarray}
where $n_\mu=\partial_\mu t$ is the unit normal vector to the hypersurfaces $t=const.$ which foliates the spacetime manifold into spatial sections $\Sigma$ and $d\Sigma=\sqrt{-g_\Sigma}dxdydz$. To find $C$ we apply the orthogonality condition $(\Phi_i(x) , \Phi_j(x)) = \delta_{ij}\delta({\bf k}_i - {\bf k}_j)$ which results in
\small
\begin{eqnarray}\label{eq47}
\begin{split}
C^2&=\frac{1}{2(2\pi)^2 \omega} \left(\int_0^l \sqrt{\frac{-g_{33}^3}{g_{00}}} Z^2(z)dz\right)^{-1},\\
&=\frac{1}{(2\pi)^2 \omega} \left((1+3\gamma_1-\gamma_0)l+(\lambda_1-\lambda_0-\frac{a}{4b})l^2
-(1+3\gamma_1-\gamma_0)P1-2(\lambda_1-\lambda_0-\frac{a}{4b})P2\right)^{-1},
\end{split}
\end{eqnarray}
\normalsize
As we show in the next section $P_1<<1$. Therefore, equation (\ref{eq47}) can be expanded as
\begin{eqnarray}\label{eq48}
\begin{split}
C^2=&\frac{1}{(2\pi)^2 \omega l} \left\{ 1+\gamma_0-3\gamma_1-(\lambda_1-\lambda_0-\frac{a}{4b})l +\left(1+\gamma_0-3\gamma_1-2(\lambda_1-\lambda_0-\frac{a}{4b})l\right)\frac{P_1}{l}      \right. \\
& \left. \;\;\;\;\;\;\;\;\;\;\;\;\;\;\;\; -2(\lambda_1-\lambda_0-\frac{a}{4b})\frac{P_2}{l}\right\},
\end{split}
\end{eqnarray}
where we have neglected second order terms like $P_1P_2, P_1^2$ and $P_2^2$.

By inserting $C^2$ into (\ref{eq44}) and (\ref{eq45}) the energy per unit area reads
\begin{equation}\label{eq49}
\begin{split}
E=
\frac{1}{2(2\pi)^2 l} \sum_\omega \int & \frac{d^2k_\perp }{\omega}
\left\{ \omega^2 \left(l-\frac{l}{2}(\lambda_1-\lambda_0) P_1+(\lambda_1-\lambda_0-\frac{a}{4b}) P_2 \right) \right. \\
 &\left. +b \left( 1+2\gamma_0-2\gamma_1-(\lambda_1-\lambda_0-\frac{a}{2b})l \right) P_1  -\frac{1}{2}\sqrt{b}(\lambda+\frac{a}{2b})P_3 \right\},
\end{split}
\end{equation}

\subsection{Computation of $P_1$,\; $P_2$ and $P_3$}
We will see in this section that $P$ parameters vanish for Dirichlet and Neumann boundary conditions while survive for Robin's one. To proceed, let's find instead $P_1$, $(\lambda_1-\lambda_0-\frac{a}{4b}) P_2$, $ (\lambda+\frac{a}{2b}) P_3$ in equation (\ref{eq49}).

\subsubsection{$P_1$}
Using expansions
\begin{eqnarray}\label{eq50}
\begin{split}
&\cos(2\sqrt{b}z+\frac{a}{2\sqrt{b}}z^2)=\cos(2\sqrt{b}z)-\frac{a}{2\sqrt{b}}z^2\sin(2\sqrt{b}z)+O(\lambda^2),\\
&\sin(2\sqrt{b}z+\frac{a}{2\sqrt{b}}z^2)=\sin(2\sqrt{b}z)+\frac{a}{2\sqrt{b}}z^2\cos(2\sqrt{b}z)+O(\lambda^2),
\end{split}
\end{eqnarray}
and after some computations, $P_1$ is given by
\begin{eqnarray}\label{eq51}
\begin{split}
\int_0^l &\cos(2\Theta)dz=\\
& \cos(2\Theta_0) \left(\frac{1}{2\sqrt{b}}(1-\frac{al}{2b})\sin(2\sqrt{b}l)-\frac{a}{8b^2}(\cos(2\sqrt{b}l)-1)+\frac{al^2}{4b}\cos(2\sqrt{b}l)\right)\\
-&\sin(2\Theta_0) \left(-\frac{1}{2\sqrt{b}}(\cos(2\sqrt{b}l)-1)+\frac{al}{4b\sqrt{b}}\cos(2\sqrt{b}l)+\frac{a}{4b}(l^2-\frac{1}{2b})\sin(2\sqrt{b}l)\right), \\
\end{split}
\end{eqnarray}
This equation can be simplified using (\ref{eq30}) or (\ref{eq34}). The result is given by
\begin{align}\label{eq52}
P_1= \frac{1}{2\sqrt{b}}(1-\frac{a l}{2b}) \mathbb{D} \cos{2\theta_0}-\frac{a l}{4b\sqrt{b}}\sin{2\theta_0},
\end{align}
where $\mathbb{D}$ stands for each of $D$ or $D^\prime$ for the limits $\kappa_1 l,\kappa_2 l<<1$ or $\beta_1 ,\beta_2 <<1$. Thus, using either (\ref{eq32})-(\ref{eq33}) or (\ref{eq36a})-(\ref{eq36b}), $P_1$ is rewritten as
\begin{subequations}
    \begin{align}[left = {P_1=\empheqlbrace}]
    &-\frac{1}{b}\left(\kappa_1+\kappa_2(1-\frac{a l}{b}) \right)  \;\;\;\;\;\;\;\;\;\ \kappa_1 l,\kappa_2 l<<1 \label{eq53a} \\ & -(\beta_1+\beta_2)     \;\;\;\;\;\;\;\;\;\;\;\;\;\;\;\;\;\;\;\;\;\;\;\;\;\;\;\;    \beta_1 ,\beta_2 <<l, \label{eq53b}
    \end{align}
\end{subequations}

\subsubsection{$(\lambda_1-\lambda_0-\frac{a}{4b}) P_2$}
First note that $(\lambda_1-\lambda_0-\frac{a}{4b}) P_2\equiv O(\lambda)P_2$.  Taking advantage of (\ref{eq50}) again and the point that $a O(\lambda)=O(\lambda^2)$ gives
\begin{eqnarray}\label{eq54}
\begin{split}
O(\lambda)P_2&=O(\lambda)\cos(2\Theta_0)\left(\frac{l}{2\sqrt{b}}\sin(2\sqrt{b}l)+\frac{1}{4b}(\cos(2\sqrt{b}l)-1)\right)\\
&-O(\lambda)\sin(2\Theta_0)\left(-\frac{l}{2\sqrt{b}}\cos(2\sqrt{b}l)+\frac{1}{4b}\sin(2\sqrt{b}l)\right)+O(\lambda^2).
\end{split}
\end{eqnarray}
A tiny calculation shows, for any $\mathbb{D}$, that $O(\lambda) \mathbb{D}\sin(2\theta_0)=O(\lambda^2)$. Employing this note and using (\ref{eq30}) or (\ref{eq34}), results in
\begin{eqnarray}\label{eq55}
O(\lambda)P_2= O(\lambda) \frac{l}{2\sqrt{b}} \biggl(\mathbb{D}\cos{2\theta_0}+\sin{2\theta_0}\biggr),
\end{eqnarray}
which, after using (\ref{eq32})-(\ref{eq33}) or (\ref{eq36a})-(\ref{eq36b}), is equivalent to
\begin{subequations}
    \begin{align}[left = {O(\lambda)P_2=\empheqlbrace}]
    &O(\lambda)\left(-\frac{l}{b}\kappa_2\right)   \;\;\;\;\;\;\;\;\;\;\;\;\;\;\; \kappa_1 l,\kappa_2 l<<1  \label{eq56a} \\ &O(\lambda)(-l\beta_2)     \;\;\;\;\;\;\;\;\;\;\;\;\;\;\;\;\;\;\;\;\;\;\;\;\;    \beta_1 ,\beta_2 <<l. \label{eq56b}
    \end{align}
\end{subequations}

Note that $P_2$ does not equal to $-\frac{l}{b}\kappa_2$ or $-l\beta_2$ individually and the above approximations are valid only in the presence of the factor $O(\lambda)$. The same point is also true for $P_3$.

\subsubsection{$ (\lambda+\frac{a}{2b}) P_3$}
The same calculation as the above shows
\begin{align}\label{eq57}
O(\lambda)P_3=O(\lambda^2),
\end{align}
where use is made of $O(\lambda) \mathbb{D}\sin(2\theta_0)=O(\lambda^2)$ again.

\subsection{summation over frequencies}
To sum over mode frequencies in equation (\ref{eq49}), we proceed as follows:
\subsubsection{$\kappa_1 l,\kappa_2 l<<1$}
Substitution of (\ref{eq53a}) and (\ref{eq56a}) into (\ref{eq49}) leads to
\begin{equation}\label{eq58}
\begin{split}
E=&\frac{1}{2(2\pi)^2} \sum_\omega \int  d^2k_\perp
\left\{ \omega \left(l+\frac{B}{2b}(\kappa_2-\kappa_1)+\frac{a}{4b^2} \kappa_2 \right) - ( 1+2A+lB )T_0 \frac{1}{\omega} \right\}\\
=&\frac{1}{2(2\pi)^2} \sum_\omega \int  d^2k_\perp
\left\{ \omega - ( 1+2A+lB )T_0 \frac{1}{\omega} \right\} \\
&+\frac{1}{2(2\pi)^2} \sum_\omega \int  d^2k_\perp B(\kappa_2-\kappa_1)\frac{ \omega}{2b}
+ \frac{1}{2(2\pi)^2} \sum_\omega \int  d^2k_\perp \frac{a\omega}{4b^2} \kappa_2.
\end{split}
\end{equation}
To find \textbf{the second summation} in the right hand side of (\ref{eq58}), we substitute $\omega$ from (\ref{eq24}) and $b$ from (\ref{eq09}). The result is given by
\begin{equation}\label{eq59}
\begin{split}
B(\kappa_2-\kappa_1)\frac{ \omega}{2b}=&B(\kappa_2-\kappa_1) \frac{ (1+A+\frac{B}{2}B )[\omega_0^2+2T_0]^\frac{1}{2} \left(1+\frac{B}{2}\frac{\kappa_2-\kappa_1}{(\frac{n\pi}{l})^2+2T_0}\right) }{2\left[(1-2A)\omega^2-k_\perp^2\right]}\\
=&\frac{B(\kappa_2-\kappa_1)}{2}     \frac{ [\omega_0^2+2T_0]^\frac{1}{2} }{\omega_0^2+2T_0-k_\perp^2}+ O(\lambda^2)+ O(\kappa^2),\\
\end{split}
\end{equation}
for which (\ref{eq22a}) has been used and $\omega_0=\sqrt{(\frac{n\pi}{l})^2+k_\perp^2}$. Let's take $y=\frac{l}{\pi}\sqrt{k_\perp^2+2T_0}$ and $a=\frac{l}{\pi}\sqrt{2T_0}$. Thus, we have
\begin{equation}\label{eq60}
\begin{split}
E_2=&\frac{1}{2(2\pi)^2} \sum_\omega \int  d^2k_\perp (\kappa_2-\kappa_1)B \frac{\omega}{2b}\\
=&\frac{(\kappa_2-\kappa_1)B}{8l}  \int_a^\infty ydy \sum_{n=0}^\infty \frac{(n^2+y^2)^\frac{1}{2}}{n^2+a^2}.
\end{split}
\end{equation}
To evaluate the sum, the Abel-Plana summation formula
\begin{eqnarray}\label{eq61}
\sum_0^\infty F(n)-\int_0^\infty F(x)dx=\frac{1}{2}F(0)+i \int_0^\infty \frac{F(ix)-F(-ix)}{e^{2\pi x}-1}dx,
\end{eqnarray}
is applicable in which $F(x)$ is defined by
\begin{eqnarray}\label{eq62}
F(x)=\frac{(x^2+y^2)^\frac{1}{2}}{x^2+a^2}.
\end{eqnarray}
The integrand in the right hand side of (\ref{eq61}) has a branch cut at $x=\pm iy$. To get ride of the branch point , it suffices, for a function of the form $F(x)=g(x^2)\sqrt{x^2+a^2}$, to have the replacement \cite{MostepanenkoBook}
\begin{eqnarray}\label{eq63}
F(ix)-F(-ix)=2i g(-x^2)\sqrt{x^2-a^2} H(x-a),
\end{eqnarray}
where $H(x-a)$ denotes the Heaviside function. Note that the integration in the left hand side of (\ref{eq61}) as well  as the constant term in the right side produces infinite terms and must be dropped. Thus, it founds that
\begin{equation}\label{eq64}
\begin{split}
E_2=&\frac{(\kappa_2-\kappa_1)B}{4l}  \int_a^\infty ydy \int_y^\infty \frac{(x^2-y^2)^\frac{1}{2}}{x^2-a^2} \frac{dx}{e^{2\pi x}-1}\\
=&\frac{(\kappa_2-\kappa_1)B}{48 \pi^2 l}  \int_{P_0}^\infty \frac{(u^2-{P_0}^2)^\frac{1}{2}}{e^u-1} du,
\end{split}
\end{equation}
where $P_0=2l\sqrt{2T_0}$. After doing the integration \cite{Gradshtyn}, we arrive at \cite{NIST}
\begin{equation}\label{eq65}
\begin{split}
E_2=&\frac{(\kappa_2-\kappa_1)B}{48 \pi^2 l}  \frac{1}{\sqrt{\pi}} (2p_0) \Gamma(\frac{3}{2}) \sum_{n=1}^\infty \frac{1}{n}K_1(p_0 n) \\
=&\frac{(\kappa_2-\kappa_1)B}{288l},
\end{split}
\end{equation}
in which $K_\nu(x)$ is the modified Bessel function and, in the last line, the approximation $K_\nu(x)\sim \frac{1}{2} \Gamma(\nu) (\frac{x}{2})^{-\nu}$ for $x<<1$ has been used.

To find \textbf{the third summation} in (\ref{eq58}) we do along the same line as the second term. First note that
\begin{equation}\label{eq66}
\begin{split}
\kappa_2\frac{ \omega a}{b^2}&=-2B\kappa_2 \frac{\omega^3}{[(1-2A)\omega^2-k_\perp^2 ]^2}
=-2B\kappa_2 \frac{(\omega_0^2+2T_0)^\frac{3}{2}}{\left[(\frac{n\pi}{l})^2+2T_0 \right]^2}+O(\kappa^2)+O(\lambda^2) \\
&=-2B\kappa_2 \frac{l}{\pi} \frac{(n^2+y^2)^\frac{3}{2}}{\left(n^2+a^2 \right)^2}\;.
\end{split}
\end{equation}
Therefore, by defining
\begin{eqnarray}\label{eq67}
F(x)=\frac{(x^2+y^2)^\frac{1}{2}}{x^2+a^2},
\end{eqnarray}
the result is found as
\begin{equation}\label{eq68}
\begin{split}
E_3=&\frac{1}{2(2\pi)^2} \sum_\omega \int  d^2k_\perp \frac{a\omega}{4b^2} \kappa_2\\
=&\frac{-B\kappa_2}{16l} \int_a^\infty ydy \sum_{n=0}^\infty \frac{(n^2+y^2)^\frac{3}{2}}{(n^2+a^2)^2}
=\frac{(\kappa_2B}{80 \pi^2 l}  \int_{P_0}^\infty \frac{(u^2-{P_0}^2)^\frac{1}{2}}{e^u-1} du \\
=&\frac{\kappa_2B}{480 l} \;.
\end{split}
\end{equation}
For the \textbf{first summation} in (\ref{eq58}), using (\ref{eq24}), (\ref{eq22a}) and (\ref{eq22b}), we observe that

\begin{equation}\label{eq69}
\begin{split}
&\omega - ( 1+2A+lB ) \frac{T_0}{\omega} \\
&=(1+A+\frac{B}{2}l)[\omega_0^2+2T_0 ]^\frac{1}{2} \left(1+B\frac{\kappa_2-\kappa_1}{(\frac{n\pi}{l})^2+2T_0}\right)^\frac{1}{2}-(1+A+\frac{B}{2}l)\frac{T_0}{\omega_0}+O(\kappa^2) \\
&=(1+A+\frac{B}{2}l) \left[ \left(\omega_0+\frac{T_0}{\omega_0}+O(T_0^2)\right) \left( 1+\frac{B}{2} \frac{\kappa_2-\kappa_1}{(\frac{n\pi}{l})^2 + 2 T_0} \right) -\frac{T_0}{\omega_0} \right] \\
&=(1+A+\frac{B}{2}l)\omega_0+ \frac{B(\kappa_2-\kappa_1)}{2}\frac{[(\frac{n\pi}{l})^2+k_\perp^2]^\frac{1}{2}}{(\frac{n\pi}{l})^2+2T_0}\;
\end{split}
\end{equation}
If summed over, the second term in the right hand side of (\ref{eq69}) equals (\ref{eq59}). The first term, except for a multiplicative constant, is the well known Casimir energy in flat spacetime. Thus
\begin{equation}\label{eq70}
\begin{split}
E_1=&(1+A+\frac{B}{2}l) \frac{1}{2(2\pi)^2} \sum_\omega \int  d^2k_\perp \omega_0+ \frac{(\kappa_2-\kappa_1)B}{480 l} \\
=&-(1+A+\frac{B}{2}l) \frac{\pi^2}{1440 l^3}+\frac{(\kappa_2-\kappa_1)B}{288 l}.
\end{split}
\end{equation}
Putting (\ref{eq65}),(\ref{eq68}) and (\ref{eq70}) altogether, the energy is given by
\begin{equation}\label{eq71}
\begin{split}
E=E_1+E_2+E_3=-(1+A+\frac{B}{2}l) \frac{\pi^2}{1440 l^3}+\frac{(\kappa_2-\kappa_1)B}{144 l}+\frac{B\kappa_2}{480 l}.
\end{split}
\end{equation}
It should be noted that we have used $\hbar=c=1$ so far.

\subsubsection{$\beta_1,\beta_2<<l$}
Using equations (\ref{eq53b}) , (\ref{eq56b}) and (\ref{eq49}) we have
\begin{equation}\label{eq72}
\begin{split}
E=&\frac{1}{2(2\pi)^2} \sum_\omega \int  d^2k_\perp
\left\{ \omega \left(1+\frac{B}{2}(\beta_1+3\beta_2)  +\frac{a }{4b} \beta_2 \right) - ( 1+2A+lB )T_0^\prime \frac{b}{\omega} \right\}\\
=&\frac{1}{2(2\pi)^2} \sum_\omega \int  d^2k_\perp \omega \left(1+\frac{B}{2}(\beta_1+3\beta_2)\right)
+\frac{1}{2(2\pi)^2} \sum_\omega \int  d^2k_\perp \frac{a \omega}{4b} \beta_2 \\
 &-\frac{1}{2(2\pi)^2} \sum_\omega \int  d^2k_\perp ( 1+2A+lB )T_0^\prime \frac{b}{\omega}.
\end{split}
\end{equation}
The mode frequencies are described now by equation (\ref{eq28}). The \textbf{first summation} on the right hand side of (\ref{eq72}) reads
\begin{equation}\label{eq73}
\begin{split}
E_4=&\left(1+\frac{B}{2}(\beta_1+3\beta_2)\right)\left(1+A+\frac{B}{2}l-\frac{B}{2}(\beta_1-\beta_2) \right)  \frac{1}{2(2\pi)^2} \sum_\omega \int  d^2k_\perp \omega_0^\prime\\
=&\left(1+A+\frac{B}{2}l+2B\beta_2\right)E_0^\prime
\end{split}
\end{equation}
in which $E_0^\prime=-\frac{\pi^2}{1440 l^{\prime 3}}$ is the Casimir energy in flat spacetime for $l^\prime =l(1+T_0^\prime)$. \\
For the \textbf{second summation} in (\ref{eq72}), we see after using (\ref{eq22a}) that
\begin{equation}\label{eq74}
\begin{split}
\beta_2\frac{a \omega}{4b}=-\frac{B\beta_2}{2} \frac{\omega^3}{\omega^2-k_\perp^2}+O(\beta^2,\lambda^2)=\frac{-B\beta_2 \pi}{2l^\prime} \frac{(n^2+y^2)^\frac{3}{2}}{n^2+\epsilon^2}, \;\; \epsilon \rightarrow 0.
\end{split}
\end{equation}
Based on this, and using the method introduced in previous sections, the second term can be shown to be
\begin{equation}\label{eq75}
\begin{split}
E_5=&\frac{-B\beta_2 \pi^2}{8l^{\prime 3}} \int_0^\infty ydy \sum_{n=0}^\infty \frac{(n^2+y^2)^\frac{3}{2}}{n^2+\epsilon^2}\\
=&\frac{B\beta_2 \pi^2}{4800 l^{\prime 3}}.
\end{split}
\end{equation}
The \textbf{third summation} in (\ref{eq72}) will be more demonstrative if written down as
\begin{equation}\label{eq76}
-(1+2A+lB )T_0^\prime \frac{b}{\omega}=-( 1+A+\frac{1}{2}B\;l )T_0^\prime \; \left( (1+Bl)\omega_0^\prime-\frac{k_\perp^2}{\omega_0^\prime}\right)+O(\beta^2,\lambda^2).
\end{equation}
Therefore, the energy $E_6$ can be found as
\begin{equation}\label{eq77}
\begin{split}
E_6=&-( 1+A+\frac{1}{2}B\;l )T_0^\prime \left( (1+Bl) E_0^\prime- \frac{ \pi^2}{4 l^{\prime 3}} \int_0^\infty y^3dy \sum_{n=0}^\infty \frac{1}{(n^2+y^2)^\frac{1}{2}} \right) \\
=&-( 1+A+\frac{1}{2}B\;l )T_0^\prime \;\left( (1+Bl) E_0^\prime- \frac{ \pi^2}{720 l^{\prime 3}} \right) \\
=&-3( 1+A+\frac{5}{6}B\;l )T_0^\prime \; E_0^\prime \;, \;\; E_0^\prime=-\frac{\pi^2}{1440 l^{\prime 3}}\;,
\end{split}
\end{equation}
where we have used the Abel-Plana formula once again. Thus, the volume energy is given by
\begin{equation}\label{eq78}
\begin{split}
E=&E_4+E_5+E_6=\left(1+A+\frac{B}{2}l+2B\beta_2\right)E_0^\prime+\frac{B\beta_2 \pi^2}{4800 {l^\prime}^3}-3( 1+A+\frac{5}{6}B\;l )T_0^\prime \; E_0^\prime
\end{split}
\end{equation}

Finally, demonstrating (\ref{eq71}) and (\ref{eq78}) in terms of the proper length $l_p$ is instructive. Thus, after some calculations we find
\small
\begin{subequations}
\begin{align}[left = {E_R=\empheqlbrace\,}\,]
    &\kappa_i l_p<<1: \nonumber \\
    &(1+\gamma_0+2\gamma_1+\frac{\lambda_0+2\lambda_1}{2}l_p) E_0+\frac{13 B}{1440 l_p}\kappa_2-\frac{10 B}{1440 l_p}\kappa_1 , \label{eq79a}\\ 
    &\beta_i <<l_p: \nonumber \\
    &(1+\gamma_0+2\gamma_1+\frac{\lambda_0+2\lambda_1}{2}l_p) E_0
    -3\left(1+\gamma_0+2\gamma_1+\frac{1}{6}(\lambda_0+8\lambda_1)l_p \right) \frac{\beta_2}{l_p}E_0 \nonumber \\ &\;\;\;\;\;\;\;\;\;\;\;\;\;\;\;\;\;\;\;\;\;\;\;\;\;\;\;\;\;\;\;\;\;\;\;\;\;\;\;\;\;\;\;\;\;\;\;\;
    -3\left(1+\gamma_0+2\gamma_1+\frac{1}{6}(5\lambda_0+4\lambda_1)l_p \right) \frac{\beta_1}{l_p}E_0 ,\label{eq79b}
\end{align}
\end{subequations}
\normalsize
in which (\ref{eq29}) and (\ref{eq22a}) have been used.

\subsection{limiting forms of the volume energy: consistency with the literature}
The results found by Saharian \cite{Saharian} \textbf{in flat spacetime}, i.e. for $\gamma_0,\gamma_1,\lambda_0,\lambda_1=0$ ,  can be recovered as follows:\\
\textbf{a)} $\kappa_1 l,\kappa_2 l<<1$: \\
Putting $\gamma_0,\gamma_1,\lambda_0,\lambda_1=0$ into equation (\ref{eq71}) produces
\begin{equation}\label{eq80}
E_0=-\frac{\pi^2}{1440 l_p^3}+O(\kappa^2)\;.
\end{equation}
which is independent of Robin coefficients! In other words, the first order correction vanishes in the flat spacetime limit. This is exactly the result which was found in \cite{Saharian}. However, note that the volume energy found in \cite{Saharian} was implicit hence we have evaluated them explicitly in Appendix A. The direct coupling between Robin coefficients and gravity in (\ref{eq71}) is interesting without which it is impossible to recover (\ref{eq80}) in the limit $\gamma_0,\gamma_1,\lambda_0,\lambda_1=0$.\\

\textbf{b)} $\beta_1,\beta_2<<l$: \\
From equation (\ref{eq78}) we find for the volume energy in flat spacetime
\begin{equation}\label{eq81}
E=(1-3T_0^\prime) \; E_0^\prime=-(1-3T_0^\prime)\frac{ \pi^2}{1440 {l^\prime}^3}=-(1-6T_0^\prime)\frac{ \pi^2}{1440 l_p^3},
\end{equation}
where we have used $l^\prime=(1+T_0^\prime)l_p$. Again we see in Appendix A that this result is consistent with the calculations done in \cite{Saharian}.

\textbf{In curved spactime}, previous studies in the literature were limited to the case of the Dirichlet or Neumann boundary conditions, i.e. $\beta_1,\beta_2,\kappa_1,\kappa_2=0$. As previously noted, in this cases, the surface energy vanishes and the volume energy equals the total energy. From (\ref{eq79a}) or (\ref{eq79b}) we find the total energy for Dirichlet or Neumann boundary conditions as
\begin{equation}\label{eq82}
E=-(1+\gamma_0+2\gamma_1+\frac{\lambda_0+2\lambda_1}{2}l_p)\frac{\pi^2}{1440 l_p^3}\equiv E_{D.N.},
\end{equation}
where $D.N.$ stands for Dirichlet or Neumann. For various spacetimes, to find $A$ and $B$, it suffices to expand the metric around $r=R$.
This result differs from the one obtained in \cite{BorzooEPJC} missing a factor of $(1+2\gamma_1+\lambda_1 l_P)$.\\
For the well-known case of \emph{Fermi coordinates}, i.e. $g_{00}=1+2gz,\; g_{ij}=\delta_{ij},\;i,j=1,2,3$, which was considered in \cite{Bimonte}-\cite{Napolitano} we see that $A=0,B=2\frac{g}{c^2}$ and the total energy equals $E=-(1+\frac{l_p}{2}g)\frac{\pi^2}{1440 l_p^3}$, in agreement with equations (5.2) in \cite{Bimonte}, (5.4) in \cite{Esposito} and (3.4) in \cite{Napolitano}.
For the case of \emph{isotropic form of the weak field limit of a typical static spacetime} (including the Schwarzschild spacetime) , i.e.
\begin{equation}\label{eq83}
ds^2=(1+2\phi)dt^2-(1-2\phi)(dx^2+dy^2+dz^2), \;\; \phi(r)<<1,
\end{equation}
in a recent paper \cite{Sorge2019}, Sorge uses the local coordinate of an observer at rest with the Casimir plates in such a way that the metric can be written as
\begin{equation}\label{eq84}
ds^2=(1+2\lambda z)dt^2-(1-2\lambda z)(dx^2+dy^2+dz^2),\;\;\; \lambda=\frac{GM}{c^2R^2},
\end{equation}
and found the energy per unit volume as
\begin{equation}\label{eq85}
\epsilon=-\frac{\pi^2}{1440l_P^4}(1-\lambda l_P).
\end{equation}
This is the special case of the result (\ref{eq82}) for $\gamma_0=\gamma_1=0,\lambda_0=-\lambda_1=\frac{GM}{c^2R^2}$. Note that $E$ was the energy per unit area hence we used $\epsilon l=E$ and $l=l_p(1+\frac{\lambda}{2}l_p)$. Therefore our calculation confirms the results found in \cite{Sorge2019}.
\section{experimental notes}
The energy formula is simple and consists of a factor $(1+\epsilon)$ times the usual Casimir energy in flat spacetime, see (\ref{eq79a})-(\ref{eq79b}) and (\ref{eq82}). Thus, the relative shift of energy is given by
\begin{equation}\label{eq86}
\frac{\delta E}{E_0}=\epsilon,
\end{equation}
where $\epsilon$ contains any information about the model by which we describe the gravity. In fact, for a theory of gravity, it suffices to find the weak field limit of the theory. i.e. the metric (\ref{eq01}), then find $\epsilon$. For Dirichlet and Neumann conditions $\epsilon=\gamma_0+2\gamma_1+\frac{\lambda_0+2\lambda_1}{2}l_p$. Therefore, the influence of any correction to the Newtonian gravity can be encoded into the constants $\lambda_0,\lambda_1,\gamma_0$ and $\gamma_1$ through the $\frac{1}{r}$-potential which appears in the weak field limit.

For a typical spacetime, say the Earth, $\epsilon\sim 10^{-10}$ near the surface of the Earth. An experimental investigation of Casimir force has been made in \cite{BorzooAnnDerPhys} using recent data obtained by employing the so-called Casimir-less technique developed by Chen et al \cite{Chen}. Fortunately, they made a significant improvment in measurement of the force between two microscopic masses at most four orders of magnitude better than existing data, i.e. $\frac{\delta E}{E_0}\sim 10^{-6}$ \cite{BorzooAnnDerPhys} . Although there are still four orders of magnitudes needed to the disired level of accuracy to detect the shift of the energy, this improvement and underlying techniques in recent researches invokes the hopes for better measurement of the Casimir force in the near future\cite{Chen,BorzooAnnDerPhys}.

\begin{figure}[!htb]
   \begin{minipage}{0.48\textwidth}
     \centering
     \includegraphics[width=0.9\linewidth]{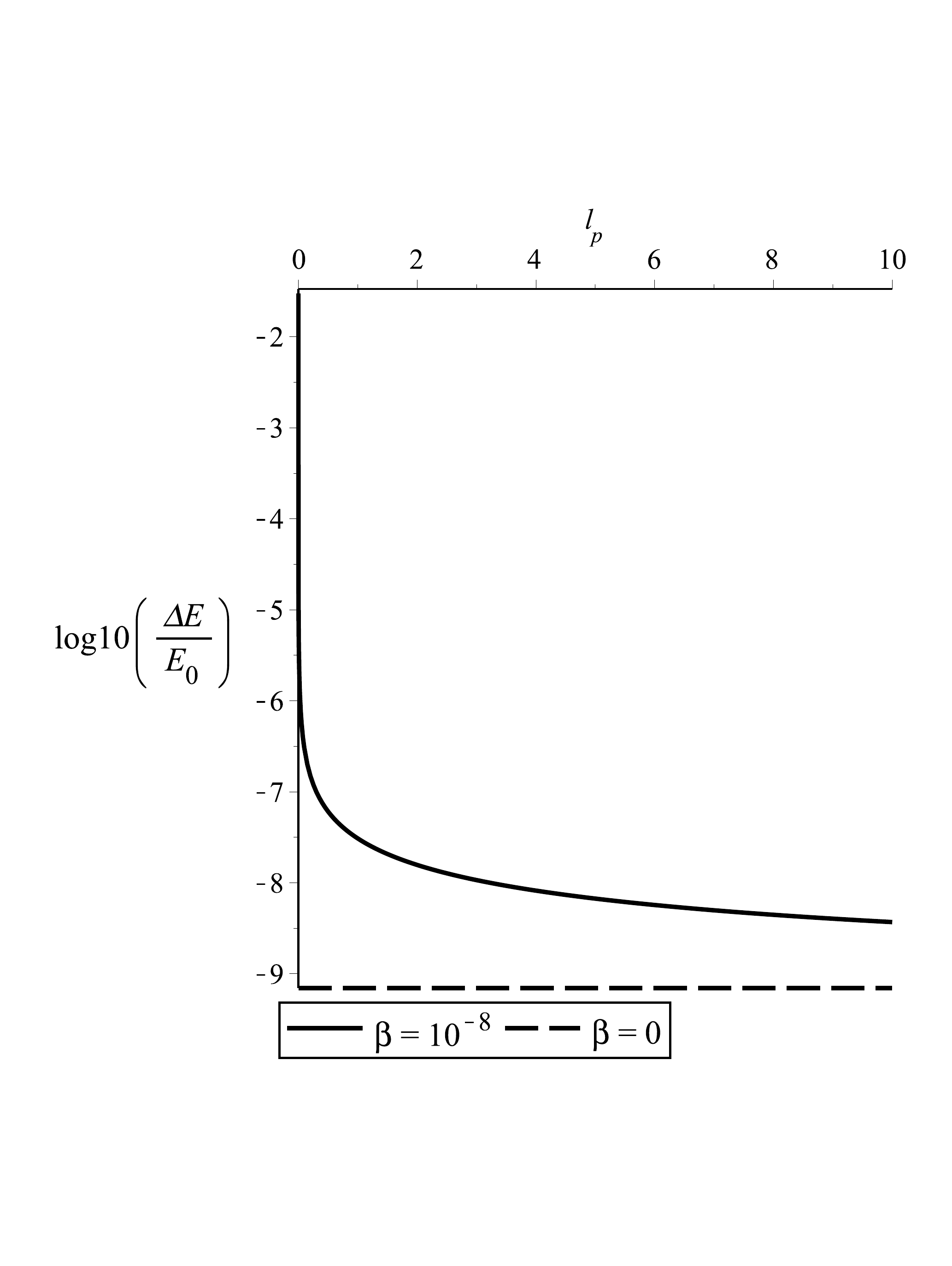}\label{fig:numerical1}
         \caption{logarithm of relative energy for $1\; \mu m<l_p<10\; m$}
   \end{minipage}\hfill
   \begin{minipage}{0.48\textwidth}
     \centering
     \includegraphics[width=0.9\linewidth]{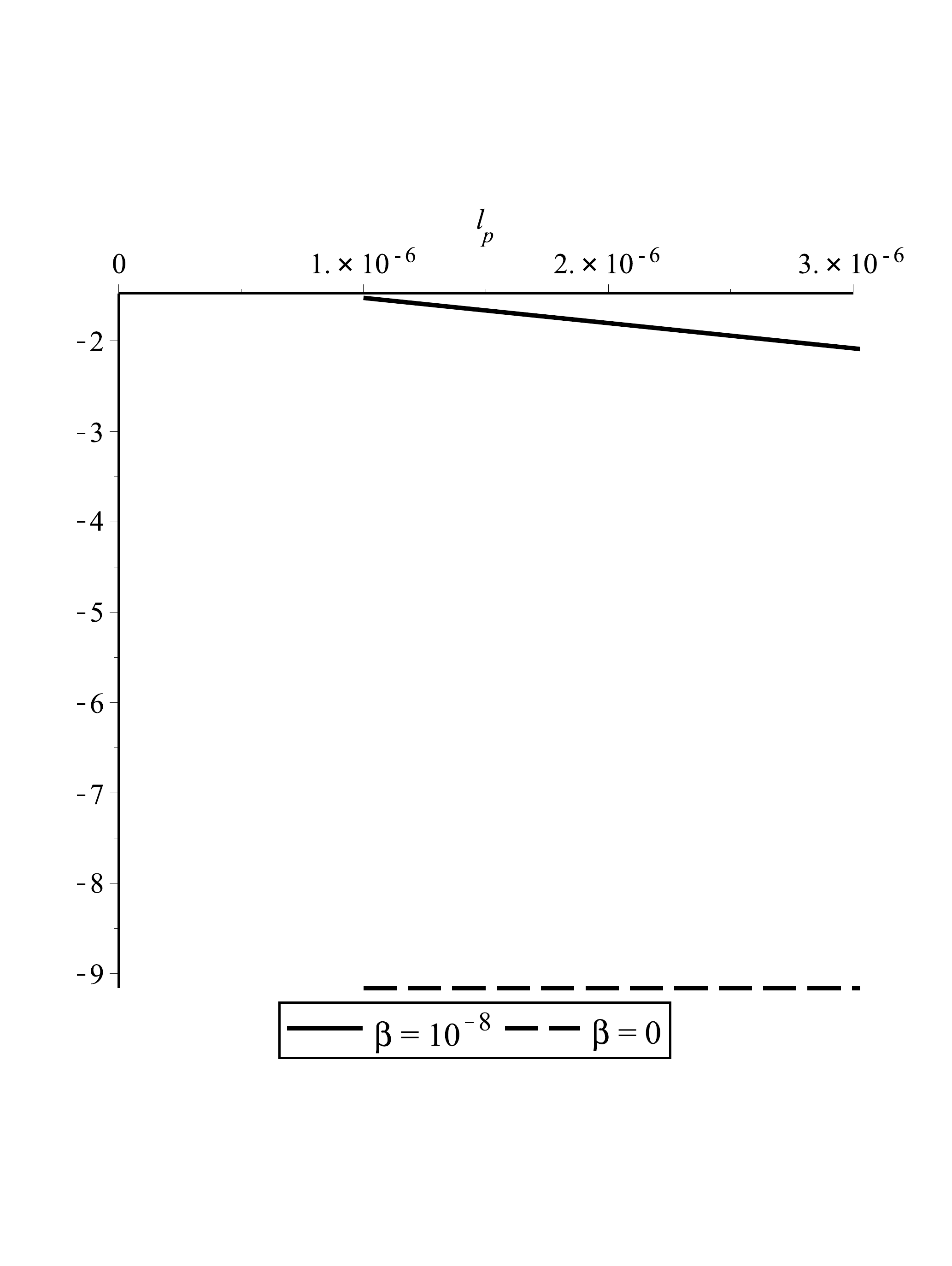}
     \caption{behavior of the logarithm of relative energy near $l_p=1\; \mu m$ }
   \end{minipage}
\end{figure}
Figure 2 shows $\log(\epsilon)$ for (\ref{eq79b}) for Shwarzschild spacetime and $\beta_2=-10^{-8},\; \beta_1=0$. Figure 3 is the zoomed version of Figure 2 for small values of $l_p$. The dashed curve describes the pure gravitational part of $\epsilon$, i.e. $\beta=\beta_2=0$ part or $E_{D.N.}$, and the solid one the whole $\epsilon$. As it is apparent, for large values of $l_p$ the energy approaches $E_{D.N.}$. This, satisfies the demand that gravity corrections grow up at large $l_p$ as well as the dominance of the corrections due to Robin coefficients in short scale. Fig.3 shows that for lowest value of $l_p$ ,which we considered, the corrections due to Robin coefficient $\beta$ approaches the value $\frac{\Delta E}{E_0}\simeq 10^{-2.5}$, which is quit remarkable compared to the corrections due to gravitational field. Using greater values of $\beta$ will shift the solid curve upward and greater contributions for $\beta$ sector of the energy can be achieved.

\section{discussion}
In this paper we analyzed the Casimir effect of parallel plates located in the curved spacetime of metric (\ref{eq01}) under Robin boundary conditions (Robin BCs). Robin BCs are generalizations to Neumann and Dirichlet ones. Our main results are equations (\ref{eq79a}) and (\ref{eq79b}) which show the influence of gravity described by metric (\ref{eq01}) on the quantum vacuum of a scalar field. Figure 2 shows the energy per distance between the plates for some typical values of Robin coefficients $\beta_2=\beta=-10^{-9}, \; \beta_1=0$. According to Figure 2, for small values of the proper distance between the plates $l_p$, the Robin coefficient has a dominant contribution to Casimir energy correction. For large values of $l_p$, the gravitational field may has as equal contribution to Casimir energy correction as Robin coefficients does.

As a special case of (79), equation (\ref{eq82}) demonstrates the energy for Neumann or Dirichlet conditions. Some special cases of (\ref{eq82}) has been found previously in the literature, i.e. for Fermi coordinates and Schwarzshild spacetime. It should be stressed that, due to the presence of $\gamma_0$ in (\ref{eq82}), the first order correction to the energy does not vanish, despite some claims in the literature. Equation (\ref{eq82}) covers all previous results found by other authors and generalizes the problem for arbitrary independent small parameters mentioned in the metric (\ref{eq01}).

Physically said, if $\gamma_0+2\gamma_1+\frac{\lambda_0+2\lambda_1}{2}l_p>0$ the energy decreases relative to that of flat spacetime (Note that $E_0<0$). Therefore, the quantum vacuum strengthen the Casimir force between the plates (and weaken the energy) in reaction to gravity. As an example, for Fermi coordinates with $\gamma_0=\gamma_1=\lambda_1=0, \lambda_0=\frac{2g}{c^2}>0$ the energy increases. For the Schwarzschild spacetime $\gamma_0=-\gamma_1=\frac{GM}{c^2 R},\lambda_0=-\lambda_1=-\frac{GM}{c^2 R^2}$ hence $\gamma_0+2\gamma_1<0$ and the energy decreases.

Using (\ref{eq22a}), equations (\ref{eq79a}) and (\ref{eq79b}) can be written in terms of $E_{D.N.}$ as follows:
\begin{subequations}
\begin{align}[left = {E_R=\empheqlbrace\,}\,]
    &E_{D.N.}+(13\kappa_2-10\kappa_1)B\frac{l_p^2}{\pi^2}E_{D.N.} \;\;\;\;\;\;\;\;\;\;\;\;\;\;\;\;\;\;\;\;\;\;\;\;\;\;\;\;\;\;\;\;\;\;\;\;\;\;\;\;\;\;\;\; \kappa_i l_p<<1,  \label{eq87a} \\ 
    &E_{D.N.}-3\left(1-\frac{1}{3}B l_p \right) \frac{\beta_2}{l_p}E_{D.N.}
    -3\left(1+\frac{1}{3}B l_p \right) \frac{\beta_1}{l_p}E_{D.N.} \;\;\;\;\;\; \beta_i <<l_p.   \label{eq87b}
\end{align}
\end{subequations}
Full consistency of (\ref{eq87a}) and (\ref{eq87b}) with the literature was shown in subsection D. Further investigation of (\ref{eq87a}) and (\ref{eq87b}) will ends up with an interesting point. Using redefinition $\kappa_2\rightarrow -\kappa_2$ and taking approximations $1+\frac{1}{3}B l_p\simeq1$ and $1-\frac{1}{3}B l_p\simeq1$ gives
\begin{subequations}
\begin{align}[left = {E_R=\empheqlbrace\,}\,]
    &E_{D.N.}-(13\kappa_2+10\kappa_1)B\frac{l_p^2}{\pi^2}E_{D.N.} \;\;\;\;\;\;\;\;\;\;\;\;\;\;\;\; \kappa_i l_p<<1,  \label{eq88a} \\ 
    &E_{D.N.}+\frac{3}{l_p}(\beta_2-\beta_1)E_{D.N.} \;\;\;\;\;\;\;\;\;\;\;\;\;\;\;\;\;\;\;\;\;\;\;\;\;\; \beta_i <<l_p.   \label{eq88b}
\end{align}
\end{subequations}
Without loss of generality suppose $B>0,\;\beta_2>\beta_1>0$. Therefore, $E<E_{D.N.}$ for $\kappa_i l_p<<1$ and $E>E_{D.N.}$ for $\beta_i<<l_p$. It turns out, according to the intermediate value theorem, that the energy $E$ must be equal to $E_{D.N.}$ somewhere between the two asymptotes $\kappa_i l_p<<1$ and $\beta_i <<l_p$ . In brief, there are some values of the Robin coefficients for which the Casimir energy equals to that of the Dirichlet and Neumann boundary conditions hence independent of the Robin coefficients!
Our last word concerns the experiment. 
The current accuracy of the experiment is not sufficient to detect the shift of energy. However, the fast progress in the development of precise measurements is hopeful and we think the shift will be measurable soon.

\appendix
\section{limiting form of the volume energy in flat spacetime}
Equation (4.31) of \cite{Saharian} represents the finite part of the volume energy in flat spacetime as $E=(\epsilon_1-2\epsilon_2)l$, where $\epsilon_1$ and $\epsilon_2$ were defined through implicit formulas
\begin{equation}\label{A1}
\begin{split}
\epsilon_1&=-\frac{1}{6\pi^2 l^4} P.V. \int_0^\infty \frac{t^3}{\frac{(b_1t-1)(b_2t-1)}{(b_1t+1)(b_2t+1)}e^{2t}-1}dt, \\
\epsilon_2&=-\frac{b_1+b_2}{6\pi^2 l^4} P.V. \int_0^\infty \frac{(1-b_1b_2t^2)t^3}{(b_1t-1)^2(b_2t-1)^2 e^{2t}-(b_1t^2-1)(b_2t^2-1)}dt,
\end{split}
\end{equation}
in which $b_1=-\frac{1}{\kappa_1},b_2=-\frac{1}{\kappa_2}$. Using \cite{BorzooAnnDerPhys}
\begin{subequations}
    \begin{align}[left = {\frac{(1-b_1t)(1-b_2t)}{(1+b_1t)(1+b_2t)}=\empheqlbrace}]
    & 1-2(b_1+b_2)t\simeq e^{-2(b_1+b_2)t} \label{A2a}  \;\;\;\;\;\;\;\;\;\ b_1,b_2<<l  \\
    & 1-\frac{2}{t}( \frac{1}{b_1}+\frac{1}{b_2} )     \;\;\;\;\;\;\;\;\;\;\;\;\;\;\;\;\;\;\;\;\;\;\;\;\;\;\;\;\;\    b_1,b_2>>l, \label{A2b}
    \end{align}
\end{subequations}
and expanding up to first order in terms of $b_1+b_2$ or $1/b_1+1/b_2$, it is straightforward to show that
\begin{subequations}
    \begin{align}[left = {\epsilon_1=\empheqlbrace}]
    &-\frac{1+4(b_1+b_2)}{6\pi^2l^4}\int_0^\infty \frac{t^3}{e^{2t}-1}dt  \;\;\;\;\;\;\;\;\;\ b_1,b_2<<l \label{A3a} \\ & -\frac{\pi^2}{1440l^4}\left(1+\frac{20}{\pi^2}(\frac{1}{b_1}+\frac{1}{b_2})\right)     \;\;\;\;\;\;\;\;\;\;\;\;\;\;\;\;\;\;\;\;\;\;\;\;\    b_1 ,b_2 >>l, \label{A3b}
    \end{align}
\end{subequations}
\begin{subequations}
    \begin{align}[left = {\epsilon_2=\empheqlbrace}]
    &\frac{b_1+b_2}{6\pi^2l^4}\int_0^\infty \frac{t^3}{e^{2t}-1}dt  \;\;\;\;\;\;\;\;\;\ b_1,b_2<<l \label{A4a} \\
    & -\frac{\pi^2}{1440l^4}\left(\frac{10}{\pi^2}(\frac{1}{b_1}+\frac{1}{b_2})\right)     \;\;\;\;\;\;\;\;\;\;\;\;\;\;\;\;\;\;\;\;\;\;\;\;\    b_1 ,b_2 >>l. \label{A4b}
    \end{align}
\end{subequations}
Using $\epsilon_1$ and $\epsilon_2$ will introduce the volume energy in equations (\ref{eq80})  and (\ref{eq81}).

\section *{Acknowledgments}
The author would like to thank University of Tehran for supporting this research under the project No. 30102/1/01 and National Science Foundation of Iran (INSF) under the priject No. 96010216.

\end{document}